\documentclass[aps,floats,twocolumn,showpacs]{revtex4}

\usepackage{pifont,colortbl,multirow,graphicx,mathrsfs,makeidx,color,epsfig,
fancyhdr,floatflt,fancybox,calc,amsmath,amsfonts,amssymb,amsthm,latexsym}

\usepackage{epsfig}
\usepackage[latin1]{inputenc}

\newcommand{\nc}{\newcommand}
\nc{\be}{\begin{equation}} \nc{\ee}{\end{equation}}
\nc{\bea}{\begin{eqnarray}} \nc{\eea}{\end{eqnarray}}
\nc{\rds}{{\rm d}s} \nc{\rdt}{{\rm d}t} \nc{\rdr}{{\rm d}r}
\nc{\rdO}{{\rm d}\Omega} \nc{\s}{{\rm S}} \nc{\Pl}{{\rm P}}
\nc{\dis}{\displaystyle} \nc{\crit}{_{\rm cr}}
\nc{\rd}{{\rm d}} \nc{\munu}{{\mu\nu}}

\begin{document}

\title{Black hole evaporation within a momentum-dependent
metric}

\author{G. Salesi}

\affiliation{Universit\`a Statale di Bergamo, Facolt\`a di
Ingegneria, viale Marconi 5, I-24044 Dalmine (BG), Italy; and
Istituto Nazionale di Fisica Nucleare, Sezione di Milano, via
Celoria 16, I-20133 Milan, Italy}

\email{salesi@unibg.it}

\author{E. Di Grezia}
\affiliation{Universit\`a Statale di Bergamo, Facolt\`a di
Ingegneria, viale Marconi 5, I-24044 Dalmine (BG), Italy; and
Istituto Nazionale di Fisica Nucleare, Sezione di Milano, via
Celoria 16, I-20133 Milan, Italy}

\email{digrezia@na.infn.it}

\

\

\begin{abstract}

\noindent We investigate the black hole thermodynamics in a ``deformed'' relativity framework
where the energy-momentum dispersion law is Lorentz-violating and the Schwarzchild-like metric is momentum-dependent with a Planckian cut-off. We obtain net deviations of the basic thermodynamical quantities from the Hawking-Bekenstein predictions: actually, the black hole evaporation is expected to quit at a nonzero critical mass value (of the order of the Planck mass), leaving a zero temperature remnant, and avoiding a spacetime singularity. Quite surprisingly, the present semiclassical corrections to black hole temperature, entropy, and heat capacity turn out to be identical to the ones obtained within some quantum approaches.


\end{abstract}

\maketitle

\section{Introduction}

\noindent Black holes are very peculiar physical systems since are fully described only recurring to classical theories, namely General Relativity and Thermodynamics, as well as to quantum gravity theories.
The merging between the above theories is required in particular for primordial and for microscopic (``mini'') black holes rather than for stellar or galactic supermassive ones: this for the extreme smallness of the Planck scale at which classical and quantum approaches appear to carry quite different previsions.

According to semiclassical Hawking-Bekenstein theory (HB) \cite{HB}, which takes into account the  quantum effects due to the very strong gravitational field on the black hole surface, we have emission of radiation out from the black hole as it be a black body at a given temperature, besides a negative-energy flux from the surface towards the interior of the black hole. As a consequence the total energy decreases in time, while the temperature increases more and more. This can be inferred from the following qualitative argument based on the Heisenberg indetermination relation. The energy of a  photon\footnote{We refer to ordinary photons: not just the ones we are going to study in this paper, for which $E\neq p$.} with a wavelength equal to the black hole radius can be assumed of the order of $1/R_\s$, $R_\s$ being the Schwarzchild radius: we then expect a temperature of the order of $1/GM$ (hereafter $\hbar = c = k_{\rm B} = 1$).
Actually, the HB calculations lead to the following relation between black hole mass and temperature
\be
T = \frac{\kappa}{2\pi} = \frac{1}{8\pi GM} \sim
10^{-7}\left(\frac{M_\odot}{M}\right)[{\rm K}]\,,
\label{temph}
\ee
where $\kappa$ is the ``surface gravitation'' and $M_\odot\sim 10^{54}$\,TeV is the solar mass.
In the HB picture the evaporation goes on until the initial mass of the black hole
is converted into radiation, and the process ends with an explosion
since, as the temperature, also the mass loss rate $dM/dt$ goes to
infinity. Besides the unphysical loss rate divergence, the total evaporation
predicted by the HB theory entails other serious problems and
inconsistencies as the baryon and lepton number non-conservation,
the ``information paradox'', and the microscopical origin
of the entropy \cite{WALD,FN}.

A complete understanding of those problems is only possible within
the framework of a quantum theory of gravity.
However we think that the effective quantum behavior can be (at least qualitatively) predicted
as well within a semiclassical theory as the one we are going to propose in the next
sections.

\section{Spacetime endowed with a momentum-dependent metric}

\noindent In recent times ultra-high energy Lorentz symmetry violations
have been investigated, both theoretically and experimentally, by means
of quite different approaches, sometimes extending, sometimes abandoning
the formal and conceptual framework of Einstein's Special Relativity.
Hereafter, for simplicity, we shall use the term ``violation'' of the Lorentz
symmetry, but in some of the below mentioned theories ---e.g., in the so-called
``Deformed'' or ``Doubly'' Special Relativity (DSR),
where ``deformed'' 4-rotation generators are considered---
although Special Relativity does not hold anymore, an underlying extended
Lorentz invariance does exist.
The most important consequence of a Lorentz violation is the
modification of the ordinary momentum-energy dipersion law
$\,E^2=p^2+m^2$, at energy scales usually assumed of the
order of the Planck energy $E_\Pl=M_\Pl c^2=\sqrt{\hbar c^5/G}$,
by means of additional terms which vanish
in the low momentum limit. Lorentz-breaking observable effects appear
in Grand-Unification Theories\cite{GUTs}, in String Theories\cite{String},
in Quantum Gravity\cite{QG}, in foam-like quantum spacetimes\cite{Foam};
in spacetimes endowed with a nontrivial topology or with a discrete structure
at the Planck length\cite{Spacetimes,Alfaro}, or with a (canonical or noncanonical)
noncommutative geometry\cite{ChenYang,GAC,NCG}; in the so-called
``extensions'' of the Standard Model incorporating breaking of Lorentz
and CPT symmetries\cite{SME}; in theories with a variable speed of light
or variable physical constants.
In particular, the M-Theory\cite{String}, the Loop Quantum Gravity
\cite{Spacetimes,Alfaro,LQG} and the Causal Dynamical Triangulation \cite{Loll}
lead to postulate an
essentially discrete and quantized spacetime, where a fundamental
mass-energy scale naturally arises, in addition to $\hbar$ and
$c$. An intrinsic length is directly correlated to the existence
of a ``cut-off'' in the transferred momentum necessary to avoid
the occurrence of ``UV catastrophes'' in Quantum Field Theories.
Moreover, some authors suspect that the Lorentz symmetry breaking
may play a role in extreme astrophysical phenomena as, e.g., the observation of
ultra-high energy cosmic rays with energies\cite{UHECR}
beyond the Greisen-Zatsepin-Kuzmin\cite{GZK} cut-off,
and of gamma rays bursts with energies beyond 20 TeV originated in distant
galactic sources \cite{Markarian}.

A natural extension of the standard dispersion law can be put in
most cases under the general form
\be
E^2 = p^2+m^2+p^2f(p/M)\,,
\label{s2}
\ee
where $M$ indicates a (large) mass scale characterizing the Lorentz violation.
By using a series expansion for $f$, under the assumption being $M$ a very large
quantity, we can consider only the lower order nonzero term in the expansion:
\be
E^2 = p^2+m^2+\alpha
p^2\left(\frac{p}{M}\right)^n\,.
\ee
The most recurring exponent in the literature on Lorentz violation is the lowest
one, i.e. $n=1$ \cite{Salesi}:
\be
E^2 =
p^2+m^2+\alpha\frac{p^3}{M}\,. \label{cubic}
\ee

\noindent An interesting theoretical approach to Lorentz simmetry violation is found
in DSR \cite{GAC,NCG,Deformed} working in $k$-deformed Lie-algebra noncommutative
($k$-Minkowski) spacetimes, in which both the Planck scale and the speed of
light act as characteristic scales of a 6-parameter group of spacetime 4-rotations
with deformed but preserved Lorentz symmetries.
In place of the ordinary constraint
$$
E^2-p^2=m^2\,
$$
in such theories the Lorentz-violating (LV) modified dispersion law can be put
in the form
\begin{equation}
E^2f^2(E)-p^2g^2(E)=m^2\,.
\label{Efpg}
\end{equation}
For example in ref.\,\cite{GAC}, where the Lie algebra is given by \ $[x_i,x_0]=i\lambda x_i$, \ $[x_i,x_k]=0$ \ ($\lambda$ being a very small length of the order of $M_\Pl^{-1}$),
the dispersion relation
\be
E^2=p^2+m^2+\lambda Ep^2
\ee
can be recovered taking
\be
f^2(E) = 1 \qquad\qquad g^2(E) = 1 + \lambda E\,.
\ee
In some DSR theories \cite{SM090,SM085,AleMa1,AleBraMa} a modified set of Special Relativity
principles is assumed: \ a) the Galileian relativity principle; \ b) the speed of light is energy-dependent, but in the small energy limit goes to the universal constant $c$ for all inertial observers; \ c) also the Planck energy-momentum is an absolute quantity, independent of the given inertial frame where is measured.
Let us quote some typical metric form factors appearing in the literature: in \cite{Ling2}
it is assumed $f=1$, $g = [1 - a(\lambda E)^n]^{-1}$; in \cite{CM}
$f=1$, $g = [1 - a(\lambda E)^n]$;
in \cite{AleMa1,SM085,SM090} $f=1$, $g = (1 + \lambda E)^\gamma$,
and in \cite{AleBraMa} $f=1$, $g = 1 + (\lambda E)^\gamma$, with $\gamma\in\Re$.
The above DSR theories have been generalized to curved spacetimes: this is the case of
the so-called ``Doubly General Relativity'', named also as ``Rainbow's Gravity'' \cite{SM2}.
The resulting metric depends on both probe energy and gravity field,
with straightforward modifications to Einstein equations for field and matter.

Let us now propose a more symmetric form of constraint (\ref{Efpg}):
\begin{equation}
E^2f^2(E)-p^2g^2(p)=m^2\,.
\label{mass-shell}
\end{equation}
We can easily see that if we make the most simple choice
\be
f^2(E) = 1 \qquad\qquad\quad g^2(p) = 1-\lambda p\,,
\label{choice}
\ee
where $\lambda\simeq M_\Pl^{-1}$ is the fundamental length scale of an underlying
LV theory, we do recover the first order (n=1) LV dispersion law (\ref{cubic}).
Notice that, because of the negative term $-\lambda p$, the energy vanishes when $p=\dis\frac{1}{\lambda}\simeq M_\Pl$. Hence quantity $1/\lambda$, in a sense,
plays the role of the ``maximal momentum'' corresponding to the ``granular''
nature of space predicted in many of the models quoted in the beginning of
this section.
Eq.\,(\ref{mass-shell}) can be equivalently written as
\be
\eta_\munu p^\mu p^\nu=m^2
\ee
with $\eta_\munu$ diagonal in flat spacetime and
\footnote{As a consequence we have also a modified commutator between space coordinates
and momenta \ $[x_i, p_k] = i\hbar\sqrt{1-\lambda p}\,\delta_{ik}\,,$ \
which vanishes for $p\lesssim 1/\lambda$, apparently implying an effective
\textit{classical} behavior in the Planckian regime.}
\be
\eta_{00} = f(E) = 1 \qquad\quad \eta_{ik} = - g(p)\,\delta_{ik} 
= - \sqrt{1 - \lambda p}\,\delta_{ik}
\label{PLmetric}
\ee
Let us stress that, at variance with other metrics adopted in literature (as the ones previously quoted), metric (\ref{PLmetric}) {\em becomes singular at the Planck momentum}.
As we shall show, the presence of a very Planckian cut-off in
the theory can overcome some of the above-mentioned problems when applying General Relativity at the Planck scale.

Just as the properties of a crystal can depend on the energy of phonons propagating in it,
analogously our spacetime geometry can depend on the moving particle energy. At low energies
the phonons cannot see the discrete structure of crystals and behave like ordinary
photons. At high energies, on the contrary, they become highly sensitive to the medium
properties and exhibit a rather exotic behavior. We therefore expect that, like phonons,
also the Hawking radiation high energy photons should behave differently
from the photons of the ordinary black body radiation.
In the next section we shall study some consequences of
metric (\ref{PLmetric}) on the thermodynamical evolution of a black hole.

\section{Modified black hole thermodynamics}

\noindent In General Relativity the metric for a non-rotating, uncharged, spherical black hole,
endowed with a mass $M$, is the Schwarzchild one:
\be \rds^2=-
\left(1-\frac{2GM}{r}\right) \rdt^2 +
\left(1-\frac{2GM}{r}\right)^{-1}\rdr^2
+r^2\rd\Omega^2\,, \label{schwarzschild0}
\ee
whilst applying Eq.\,(\ref{PLmetric}) we have the following modified metric \cite{SM2}
$$ 
\rds^2=- \frac{1}{f^2(E)}\left(1-\frac{2GM}{r}\right)\rdt^2 +
$$
\be
+ \frac{1}{g^2(p)}\left[\left(1-\frac{2GM}{r}\right)^{-1}\rdr^2
 + r^2\rd\Omega^2\right]\,. 
\ee
As abovesaid, according to the HB theory the black hole temperature can be taken equal to $\dis\frac{\kappa}{2\pi}$, where $\kappa$ is the surface gravity.
Because of the chosen metric, the black hole surface gravity and temperature do depend on the probe energy:
\be
\kappa=-\lim_{r\to R_\s}\sqrt{\frac{-g^{rr}}{g^{tt}}}\frac{(g^{tt})'}{g^{rr}}
=\frac{g(p)}{f(E)}\frac{1}{4GM}\,,\label{surfgrav}
\ee
\be
T= \frac{\kappa}{2\pi}=\frac{g(p)}{f(E)}\frac{1}{8\pi GM}\,.\label{temp1}
\ee
Hence, with our choice for $f(E)$ and $g(p)$, we get
\be
T= \frac{\sqrt{1-\lambda p}}{8\pi GM}\,.
\label{temperatura}
\ee
\noindent Let us apply the ordinary uncertainty relation to photons near the event horizon:
\be
p\simeq{\rm\delta} p\sim\frac{1}{{\delta}
x}\sim \frac{1}{4\pi R_s}= \frac{1}{8\pi GM}\,.
\ee
By inserting the previous equation in (\ref{temperatura}) we find
\be
T=\frac{\sqrt{1- \lambda p}}{8\pi GM}
\sim\frac{\dis\sqrt{1-\frac{\lambda}{8\pi GM}}}{8\pi GM}\,.\label{temp3}
\ee
Being $p\leq 1/\lambda$ we shall consider only $M\geq M\crit\equiv\dis\frac{\lambda}{8\pi G}$.
\begin{figure}
\begin{center}
\epsfig{height=5truecm,file=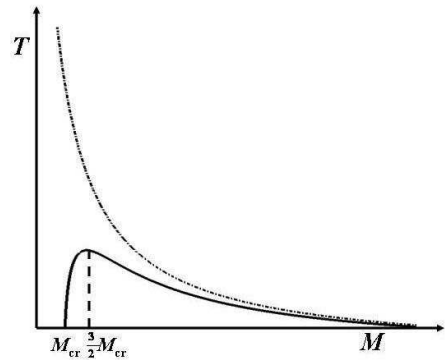}
\caption{Temperature as a function of the mass (compared with the standard HB plot)}
\label{fig1}
\end{center}
\end{figure}

\noindent For $M\gg M\crit$ we can approximate the black hole temperature as follows
\be
T\sim\frac{1}{8\pi GM} \left[1-\frac{1}{2}\frac{M\crit}{M}-
\frac{1}{8}\left(\frac{M\crit}{M}\right)^2 +{\rm O}\left(\frac{M\crit^3}{M^3}\right)\right]\,,
\label{exptemp3}
\ee
while, when $M$ approaches $M\crit$, we can assume
\be
T\sim\frac{1}{8\pi
GM\crit^{3\over 2}}\left[\sqrt{M-M\crit} + M\crit^{-{1\over 2}}{\rm O}(M-M\crit)\right]\,. \label{exptempmcr}
\ee
Therefore the black hole temperature, by contrast with the HB theory, does not diverge at $M=0$, but has a finite maximum at $\dis M=\frac{3}{2}M\crit$ (Fig.\ref{fig1}).

Assuming that the first principle of thermodynamics is still valid, namely $\rd Q= \rd M = T\rd S$, we can obtain the intrinsic entropy by inserting the temperature (\ref{temp1}) into this relation, and then integrating
\begin{eqnarray} 
&&S = \int\frac{\rd M}{T} \sim \int\frac{8\pi G M}{\sqrt{1-\frac{M\crit}{M}}}\,\rd M =\nonumber
\\ &&
= \frac{2\pi G}{\sqrt{1-\frac{M\crit}{M}}}\left(2M^2 + MM\crit -
3M\crit^2\right. + \nonumber\\&&   
\left.+3M\crit^2\sqrt{1 - \frac{M\crit}{M}}\,{\rm arcsinh}\sqrt{\frac{M}{M\crit}-1}\,\right) 
\label{entropia}
\end{eqnarray}

\begin{figure}
\begin{center}
\epsfig{height=5truecm,file=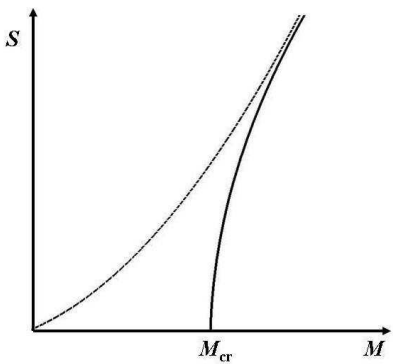}
\caption{Entropy as a function of the mass
(compared with the standard HB plot)}
\label{fig2}
\end{center}
\end{figure}

\noindent Of course, for $\lambda\to 0$ we recover the classical HB result
\be
\lim_{\lambda\to 0}S = \frac{1}{4}A\,,
\ee
where $A$ indicates the horizon area $16\pi GM^2$. The same classical behavior (quadratic
in $M$), is obtained for very heavy black holes, i.e. \!\!\!\!\! for $M\gg M\crit$.
Expanding Eq.\,(\ref{entropia}) for $M\sim M\crit$ we obtain (see Fig.\ref{fig2})
\be
S\sim 4\pi GM\crit^{3/2}\sqrt{M-M\crit}\,. \label{expentropmcr}
\ee
Then the black hole entropy reaches its minimum (zero) together with the black hole mass,
i.e. for the critical value $M\crit$.

For the sake of comparison let us recall that, still starting from modified dispersion laws and generalized Heisenberg uncertainty relations, some authors \cite{GACnew} obtain in the entropy formula a term logarithmically dependent on the black hole mass, a result found also in some String Theory and Loop Quantum Gravity computations.

\begin{figure}
\begin{center}
\epsfig{height=5truecm,file=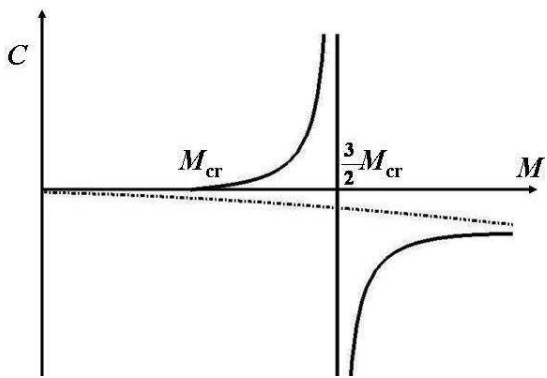}
\caption{Heat capacity as a function of the mass (compared with the standard HB plot)}
\label{fig3}
\end{center}
\end{figure}

Finally, let us calculate the black hole heat capacity. For $M\sim M\crit$ we get
\be
C =T\frac{\partial S}{\partial T}= \left(\frac{\partial T}{\partial M}\right)^{-1} \sim
16\pi G\frac{M^{5/2}\sqrt{M-M\crit}}{3M\crit -2M}
\ee
Notice (Fig.\ref{fig3}) that the heat capacity diverges at $M=\frac{3}{2}M\crit$,
corresponding to the maximum black hole temperature, and vanishes at the minimum mass $M=M\crit$.

\section{Conclusions}

\noindent We have studied the black hole evaporation applying a momentum-dependent metric,
corresponding to the lowest-order ($\sim p^3$) extension of the ordinary energy-momentum law dispersion. The HB inverse proportionality relation between mass and temperature is recovered only
for early stages of the evaporation process; whilst in the final stage mass and temperature decrease together so that at the end we have a cold ``extremal'' black hole endowed with a critical mass $M=M\crit$ of the order of the Planck mass. Correspondingly the black hole entropy reaches its minimum. At variance with the standard previsions, we also find that at $M=\frac{3}{2}M\crit$
the temperature reaches a finite maximum (of the order of the Planck temperature) and the heat capacity
diverges.

Some recent theoretical \cite{BR,BR1,FIS}  pictures of the black hole thermodynamics and evaporation which take into account quantum effects expected when the energy scale approaches the Planck energy
(that is soon before the total collapse), result in avoiding undesired divergences and spacetime singularities. In \cite{BR1} Bonanno and Reuter study the quantum gravity effects for a spherical black hole assuming a ``running'' Newton constant $G = G(k(r))$ obtained from the evolution for the effective scale-dependent gravitational action, by means of exact renormalization group equations. Using the ``Einstein-Hilbert truncation method'' they find an exact, non-perturbative solution to the evolution equation for $G(k)$.
Actually,  \textit{the quantum computations performed by those authors and the present semiclassical analysis lead to same physical predictions}. As a matter of fact, the behaviors of temperature, entropy and heat capacity as functions of the mass obtained in ref.\,\cite{BR1}
result to be (also analytically) identical to the ones found in this paper, then entailing a nonzero-mass zero-entropy $T=0$ remnant as well.
From direct comparison between ref.\,\cite{BR1} and the present theory we derive, as expected, that
our mass scale $\lambda$ is just of the order of the Planck mass.

We can conclude that, inside a classical noncommutative spacetime scenario which, as discussed in Section II, appears physically plausible on the ground of various theoretical and experimental arguments, we have obtained a reliable picture of the black hole thermodynamics which overcomes some unphysical features of the HB theory: thus encouraging further theoretical studies in this direction.

\vspace{0.5cm}

\noindent {\large{\bf Acknowledgments}}

\vspace*{0.2cm}

\noindent We are glad to thank A. Bonanno and S. Esposito for useful
hints and stimulating discussions.

\end{document}